\documentclass[aps,pre,twocolumn,superscriptaddress,showpacs,a4paper]{revtex4-1}

\usepackage[utf8]{inputenc}

\usepackage{microtype}

\usepackage{amsfonts}
\usepackage{amsmath}
\usepackage{amssymb}

\usepackage{graphicx}
\usepackage{dcolumn}
\usepackage{hyperref}
\usepackage{natbib}

\newcommand{\vect}[1]{{\mathbf{#1}}}
\newcommand{\ind}[1]{{\text{#1}}}

\let\Re\undefined
\let\Im\undefined
\DeclareMathOperator{\Re}{Re}
\DeclareMathOperator{\Im}{Im}

\begin{document}

\title{Clustering of inelastic soft spheres in homogeneous turbulence}

\author{Thomas Burgener}
\email{buthomas@ethz.ch}
\affiliation{Computational Physics, IfB, ETH-H\"onggerberg, Schafmattstrasse 6, 8093 Z\"urich, Switzerland}
\author{Dirk Kadau}
\email{dkadau@ethz.ch}
\affiliation{Computational Physics, IfB, ETH-H\"onggerberg, Schafmattstrasse 6, 8093 Z\"urich, Switzerland}
\author{Hans J. Herrmann}
\email{hans@ifb.baug.ethz.ch}
\affiliation{Computational Physics, IfB, ETH-H\"onggerberg, Schafmattstrasse 6, 8093 Z\"urich, Switzerland}
\affiliation{Departamento de Fis\'ica, Universidade Federal do Cear\'a, Campus do Pici, 60451-970 Fortaleza, Cear\'a, Brazil}

\date{\today}

\begin{abstract}
  In this paper we numerically investigate the influence of dissipation during particle collisions in an homogeneous turbulent velocity field by coupling a discrete element method to a Lattice-Boltzmann simulation with spectral forcing. We show that even at moderate particle volume fractions the influence of dissipative collisions is important. We also investigate the transition from a regime where the turbulent velocity field significantly influences the spatial distribution of particles to a regime where the distribution is mainly influenced by particle collisions.
\end{abstract}


\pacs{47.27.T-,47.27.E-,83.10.Rs,45.70.-n}

\maketitle

\section{Introduction}
Particles suspended in a fluid can show rather complex behavior and an accurate description of these effects is a longstanding challenge. In homogeneous isotropic turbulence preferential concentration \cite{Eaton1994,Fessler1994,Bec2007,Balachandar2010} is a well known effect. There small, heavy particles tend to concentrate in regions where the vorticity of the fluid velocity field is low and the strain is high. This effect is most pronounced when the Stokes number $\mathrm{St}$ of the particles is $\mathrm{St}\approx 1$. In the case of turbulence the Stokes number \cite{Collins2004} depends on the particle and fluid densities $\rho_{\ind{p}}$ and $\rho_{\ind{f}}$ as well as the particle radius $R_{\ind{p}}$ and the Kolmogorov length scale $\eta$:
\begin{equation}
  \mathrm{St}=\frac{4}{18}\frac{\rho_{\ind{p}}}{\rho_{\ind{f}}}\left(\frac{R_{\ind{p}}}{\eta}\right)^{2}.
  \label{eqn:StokesNumber}
\end{equation}

Preferential concentration has been investigated intensively by experiments \cite{Eaton1994,Fessler1994} and numerical simulations \cite{Cencini2006,Bec2005,Bec2006,Biferale2006}. The occurrence of preferential concentration is closely related to the difference in inertia between the particles and the fluid, which is expressed by the Stokes number~\eqref{eqn:StokesNumber}. For small $\mathrm{St}$ particles follow the fluid stream lines rather closely and therefore, at least in an incompressible fluid, initially homogeneous distributed particles remain distributed that way. For large Stokes numbers the particle motion is only weakly influenced by the fluid which then leads to a diffusion-like motion of the particles. For intermediate Stokes numbers the local structure of the fluid velocity field becomes important \cite{Bec2005}. In the case of dilute particle suspensions, when collisions between individual particles can be ignored, heavy particles tend to concentrate in regions of high strain rate and low vorticity, and light particles tend to do the opposite \cite{Balachandar2010}. The mechanism behind this process is that the vortices present in a turbulent flow act as centrifuges that eject heavy particles and attract light ones \cite{Bec2007}. This effect is most pronounced for $\mathrm{St}\approx 1$ \cite{Eaton1994,Fessler1994,Balachandar2010}. In most simulations one concentrates on an accurate description of the fluid and particle collisions are either ignored completely or treated as perfectly elastic. But in fact inelastic collisions are another mechanism that can lead to a clustering of particles -- an effect known as collisional cooling. As shown by Luding \textit{et al.} \cite{Luding1999,Miller2004} for freely moving particles, this effect of (free) cooling is already important for moderate amounts of dissipation.

In this paper we want to investigate the influence of dissipative collisions on the clustering of soft spheres in homogeneous turbulence. Additionally the dependence on the particle volume fraction shall be considered. We start by presenting the coupling of a discrete element model to a Lattice-Boltzmann simulation where a spectral forcing technique \cite{Alvelius1999,Cate2006} is used to generate turbulence. We then investigate the influence of the dissipation during collision on the clustering of particles for different Stokes numbers. Finally we vary the volume fraction by changing the number of particles in our system and study its effects.

\section{Model description}\label{sec:model}
The discrete element model (DEM) \cite{Cundall1979} is a widely used and well established method for simulating granular materials \cite{Herrmann1998,Luding1998,Luding2008}. We use this model here to evolve a set of $n_{\ind{p}}$ spherical particles with positions $\vect{x}_{i}$, masses $m_{i}$ and radii $R_{i}$ according to Newtons' equation of motion
\begin{equation}
  m_{i}\frac{d^{2}}{dt^{2}}\vect{x}_{i} = \vect{F}_{i},
  \label{eqn:NewtonEquation}
\end{equation}
where $\vect{F}_{i}$ is the total force acting on particle $i$. This force is given by the sum of a collision force and a drag force due to a fluid
\begin{equation}
  \vect{F}_{i}=\vect{F}_{i}^{\ind{coll}} + \vect{F}_{i}^{\ind{drag}}.
  \label{eqn:F=Fcoll+Fdrag}
\end{equation}
The collision force is given by a sum of two-particle collisions
\begin{equation}
  \vect{F}_{i}^{\ind{coll}} = \sum_{i\neq j}\vect{F}_{ij}^{\ind{coll}}
  \label{eqn:Fi=sumFij}
\end{equation}
and these two-particle collisions are modeled by a linear spring dash-pot model. In this model one first calculates the overlap between two particles $i$ and $j$
\begin{equation}
  \delta_{ij}=\left( R_{i}+R_{j} \right) - \lvert \vect{x}_{i}-\vect{x}_{j} \rvert
  \label{eqn:overlap}
\end{equation}
and if $\delta_{ij}$ is positive, a repulsive dissipative force
\begin{equation}
  \vect{F}_{ij}^{\ind{coll}}=\left( k_{\ind{n}}\delta_{ij}+c_{\ind{n}}\dot{\delta}_{ij} \right)\vect{n}_{ij}
  \label{eqn:F_ij_coll}
\end{equation}
in the direction
\begin{equation}
  \vect{n}_{ij}=\frac{\vect{x}_{i}-\vect{x}_{j}}{\lvert \vect{x}_{i}-\vect{x}_{j} \rvert}
  \label{eqn:n_ij}
\end{equation}
acts on particle $i$. The model corresponds to a damped linear spring with stiffness $k_{\ind{n}}$ and damping coefficient $c_{\ind{n}}$. The value $\dot{\delta}_{ij}$ is given by the relative velocities of the two particles in the direction of $\vect{n}_{ij}$, i.e.
\begin{equation}
  \dot{\delta}_{ij}=-\left( \vect{v}_{i}-\vect{v}_{j} \right)\cdot\vect{n}_{ij},
  \label{eqn:delta_ij_dot}
\end{equation}
where $\vect{v}_{i}$ is the velocity of particle $i$. Instead of the damping coefficient $c_{\ind{n}}$ it may be more convenient to work with the coefficient of restitution $e_{\ind{n}}$, which is an easier to handle material parameter. This coefficient is a measure for how much energy is retained after a collision. To relate $c_{\ind{n}}$ with $e_{\ind{n}}$ we additionally introduce the collision time $t_{\ind{c}}$ as
\begin{equation}
  t_{\ind{c}}=\frac{1}{\omega_{\ind{c}}}\left( \pi -2\arctan \frac{\eta_{\ind{c}}}{\omega_{\ind{c}}} \right),
  \label{eqn:tc}
\end{equation}
where
\begin{equation}
  \omega_{\ind{c}}^{2}=\frac{k_{\ind{n}}}{m_{ij}}-\frac{c_{\ind{n}}^{2}}{4m_{ij}^{2}}\qquad\text{and}\qquad \eta_{\ind{c}}=\frac{c_{\ind{n}}}{2m_{ij}},
  \label{eqn:omega_c,eta_c}
\end{equation}
with
\begin{equation}
  m_{ij}=\frac{m_{i}m_{j}}{m_{i}+m_{j}}.
  \label{eqn:m_ij}
\end{equation}
The coefficient of restitution is then given by
\begin{equation}
  e_{\ind{n}}=e^{-\eta_{\ind{c}}t_{\ind{c}}}.
  \label{eqn:en}
\end{equation}

For the drag force $\vect{F}_{i}^{\ind{drag}}$ we use an empirical drag law. For a laminar flow the drag force \cite{Bini2007} should be proportional to the difference between particle velocity $\vect{v}_{i}$ and fluid velocity $\vect{u}$ 
\begin{equation}
  \vect{F}_{i}^{\ind{drag}} = \frac{\vect{u}-\vect{v}_{i}}{\tau_{i}},
  \label{eqn:DragStokes}
\end{equation}
where $\tau_{i}$ is the particle response time
\begin{equation}
  \tau_{i}=\frac{4}{18}\frac{R_{i}^{2}}{\nu_{\ind{f}}}\frac{\rho_{i}}{\rho_{\ind{f}}}.
  \label{eqn:particleResponseTime}
\end{equation}
Here $\nu_{\ind{f}}$ is the kinematic viscosity of the fluid. For a turbulent flow on the other hand the drag force should quadratically depend on the velocity difference. A widely used and well established empirical drag law for the turbulent case \cite{Li2003,Zhu2007,Bini2007} is given by
\begin{equation}
  \vect{F}_{i}^{\text{drag}} = m_{i}\frac{3}{8}\frac{C_{\ind{D}}}{R_{i}}\left( \frac{\rho_{\ind{f}}}{\rho_{i}} \right) \lvert \vect{u}-\vect{v}_{i} \rvert  \left( \vect{u}-\vect{v}_{i} \right),
  \label{eqn:DEM_Fdrag}
\end{equation}
where the drag coefficient $C_{\ind{D}}$ is given by
\begin{equation}
  C_{\ind{D}} = \begin{cases}
    \frac{24}{\mathrm{Re}_{\ind{p}}}\left( 1+0.15\ \mathrm{Re}_{\ind{p}}^{0.687} \right) & \mathrm{Re}_{\ind{p}}<1000\\
    0.44 & \mathrm{Re}_{\ind{p}}\geq 1000
  \end{cases}.
  \label{eqn:DEM_CD}
\end{equation}
This coefficient depends on the particle Reynolds number $\mathrm{Re}_{\ind{p}}$, which is given by
\begin{equation}
  \mathrm{Re}_{\ind{p}} = \frac{2R_{i}\lvert \vect{u}-\vect{v}_{i} \rvert}{\nu_{\ind{f}}}.
  \label{eqn:DEM_ReynoldsParticle}
\end{equation}
Such a drag law describes a simplified dynamics for the particles, where the added mass effect as well as the Basset–Boussinesq history force are neglected. Such an assumption is reasonable for heavy particles. There is also no pressure gradient term present, since the fluid field is incompressible.

\begin{figure}[!tbp]
  \includegraphics{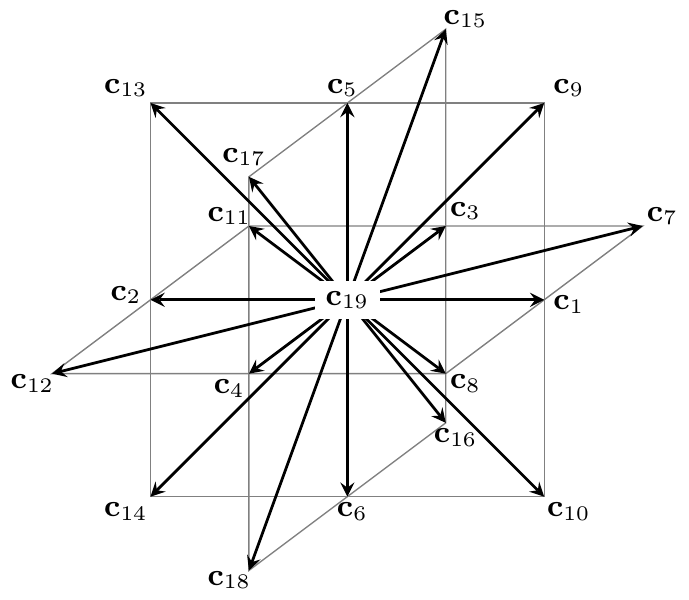}
  \caption{Definition of the lattice velocity vectors $\vect{c}_{1},\ldots,\vect{c}_{19}$ for the D3Q19 lattice.}
  \label{fig:D3Q19Def}
\end{figure}

To calculate the fluid velocity $\vect{u}$ we use a Lattice-Boltzmann method with spectral forcing \cite{Alvelius1999,Cate2006}. In recent years the Lattice-Boltzmann (LB) method has become a very successful method for solving many different problems in fluid dynamics \cite{Aidun2010,Mendoza2010,Mendoza2011}. The standard LB equation with an external force $\vect{g}\left( \vect{x},t \right)$ reads
\begin{multline}
  f_{i}\left( \vect{x}+\vect{c}_{i}\,\delta t,t+\delta t \right) - f_{i}\left( \vect{x},t \right) =\\
  -\omega\,\delta t\bigl( f_{i}\left( \vect{x},t \right) - f_{i}^{\ind{eq}}\left( \vect{x},t \right)  \bigr) + G_{i}\left( \vect{x},t \right),
  \label{eqn:LBStandart}
\end{multline}
where $f_{i}\left( \vect{x},t \right)$ is the probability of finding a particle at time $t$ and lattice site $\vect{x}$, that is moving in the direction of the discrete lattice velocity $\vect{c}_{i}$. Here $i=1,\dots ,Q$ indexes the discretization of velocity space. Equation \eqref{eqn:LBStandart} is a discretized version of the Boltzmann-equation where the left-hand side relates to the free-streaming and the right-hand side is an approximation of the collision operator plus a forcing term $G_{i}$. Here the well established Bhatnagar–Gross–Krook relation \cite{Bhatnagar1954} was used, which models a simple relaxation of the population $f_{i}$ toward a local equilibrium $f_{i}^{\ind{eq}}$ with relaxation frequency $\omega$. The equilibrium distributions are given by a Maxwell-Boltzmann distribution which is expanded up to second order in terms of Hermite polynomials resulting in
\begin{equation}
  f_{i}^{\ind{eq}} = \omega_{i}\rho_{\ind{f}}\left( 1 + \frac{\vect{c}_{i}\cdot\vect{u}}{c_{\ind{s}}} + \frac{\left( \vect{c}_{i}\cdot\vect{u} \right)^{2}}{2c_{\ind{s}}^{4}} - \frac{\lvert \vect{u} \rvert^{2}}{2c_{\ind{s}}^{2}} \right),
  \label{eqn:LB_feq}
\end{equation}
where $\omega_{i}$ are lattice weights and $c_{\ind{s}}$ is the speed of sound. In this paper we chose a D3Q19 lattice depicted in Fig.~\ref{fig:D3Q19Def}.
The lattice weights for this lattice are given as
\begin{equation}
  w_{i}=
  \begin{cases}
    \frac{2}{36} & i=1,\ldots,6\\
    \frac{1}{36} & i=7,\ldots,18\\
    \frac{12}{36} & i=19
  \end{cases}
  \label{eqn:LB_D3Q19weights}
\end{equation}
and the speed of sound is
\begin{equation}
  c_{\ind{s}}=\frac{1}{\sqrt{3}}.
  \label{D3Q19speedOfSound}
\end{equation}
The fluid density $\rho_{\ind{f}}$ and the velocity $\vect{u}$ are related to the probability distributions $f_{i}$ through \cite{Guo2002}
\begin{gather}
  \rho_{\ind{f}} = \sum_{i} f_{i}\\
  \rho_{\ind{f}} \vect{u} = \sum_{i} f_{i}\vect{c}_{i} + \frac{\delta t}{2}\vect{g}.
  \label{eqn:LB_rho_u}
\end{gather}
The choice for the forcing terms $G_{i}$ is not trivial, since one has to construct $Q$ values $G_{i}$ from a three-dimensional vector $\vect{g}$. The choice should guarantee, that the correct incompressible Navier-Stokes equations are recovered when performing the Chapman-Enskog expansion. Here we are using the expression by Guo \textit{et al.} \cite{Guo2002}
\begin{equation}
  G_{i}=\left( 1-\frac{1}{2}\omega \right)\omega_{i}\left( \frac{\vect{c}_{i}-\vect{u}}{c_{\ind{s}}^{2}} + \frac{\left( \vect{c}_{i}\cdot\vect{u} \right)}{c_{\ind{s}}^{4}}\,\vect{c}_{i} \right)\cdot\vect{g}.
  \label{eqn:LB_forcingTerms}
\end{equation}

Finally we have to specify the external force $\vect{g}$, which is the driving mechanism for the fluid. This force has to be chosen such that an homogeneous and isotropic turbulent velocity field is generated. This task is frequently encountered in direct numerical simulations (DNS) and other cases (see e.g. \cite{Eswaran1988,Alvelius1999} and references therein). In this work we use a method introduced by Alvelius \cite{Alvelius1999} which has already been used with the LB method by ten Cate \textit{et al.} \cite{Cate2006}. Here only a brief review of the technique is given.

The basic idea for calculating $\vect{g}\left( \vect{x},t \right)$ is to generate in Fourier space a random, divergence free force field $\hat{\vect{g}}\left( \vect{k},t \right)$ that is only active at small wave-vectors $\vect{k}$, and then take the inverse Fourier transform to get the force in real space. This corresponds to the picture of turbulence where energy is injected into the system at small $\vect{k}$, which is than transported to higher and higher frequencies until it is finally dissipated. To fulfill the condition
\begin{equation}
  \vect{k}\cdot\hat{\vect{g}}=0
  \label{eqn:kdotg=0}
\end{equation}
we have to choose a force $\hat{\vect{g}}$ which is always perpendicular to $\vect{k}$. This can be achieved by choosing
\begin{equation}
  \hat{\vect{g}}\left( \vect{k},t \right) = A_{\ind{ran}}\left( \vect{k},t \right)\vect{e}_{1}\left( \vect{k} \right) + B_{\ind{ran}}\left( \vect{k},t \right)\vect{e}_{2}\left( \vect{k} \right),
  \label{eqn:LB_g_hat_def}
\end{equation}
where $A_{\ind{ran}}$ and $B_{\ind{ran}}$ are two random amplitudes and the unit vectors $\vect{e}_{1}$ and $\vect{e}_{2}$ are chosen perpendicular to $\vect{k}$ and each other. Alvelius \cite{Alvelius1999} choice for these vectors is
\begin{subequations}\label{eqn:LB_e1e2}
\begin{equation}
  \vect{e}_{1}=\frac{1}{\sqrt{k_{x}^{2}+k_{y}^{2}}}
  \begin{pmatrix}
    k_{y}\\
    -k_{x}\\
    0
  \end{pmatrix}
  \label{eqn:LB_e1}
\end{equation}
and
\begin{equation}
  \vect{e}_{2}=\frac{1}{k\sqrt{k_{x}^{2}+k_{y}^{2}}}
  \begin{pmatrix}
    k_{x}k_{z}\\
    k_{y}k_{z}\\
    -\left( k_{x}^{2}+k_{y}^{2} \right)
  \end{pmatrix},
  \label{eqn:LB_e2}
\end{equation}
\end{subequations}
where $k=\lvert \vect{k} \rvert$. The amplitudes $A_{\ind{ran}}$ and $B_{\ind{ran}}$ are chosen as
\begin{subequations}\label{eqn:LB_ArandBrand}
\begin{align}
  A_{\ind{rand}}& = \sqrt{\frac{S(k)}{2\pi k^{2}}}\exp\left( i\theta_{1} \right)\sin\left( \phi \right)\\
  B_{\ind{rand}}& = \sqrt{\frac{S(k)}{2\pi k^{2}}}\exp\left( i\theta_{2} \right)\cos\left( \phi \right),
\end{align}
\end{subequations}
where $\phi,\theta_{1},\theta_{2}\in\left[0,2\pi\right]$ are chosen at every time step as uniformly distributed random numbers. The spectrum function $S\left( k \right)$ should only be active in an interval $\left[ k_{\ind{a}},k_{\ind{b}} \right]$ at small wave number and is given by a Gaussian
\begin{equation}
  S\left( k \right)=
  \begin{cases}
    A\exp\left( -\frac{\left( k-k_{\ind{f}} \right)^{2}}{c} \right) & k\in\left[ k_{\ind{a}},k_{\ind{b}} \right]\\
    0 & \text{otherwise}
  \end{cases}
  \label{eqn:LB_FkGaussian}
\end{equation}
where $c$ determines the width of $S\left( k \right)$, $k_{\ind{f}}$ the position of its maximum, and
\begin{equation}
  A = \frac{P_{\ind{in}}}{\Delta t}\frac{1}{\int_{k_{\ind{a}}}^{k_{\ind{b}}}\exp\left( -\frac{\left( k-k_{\ind{f}} \right)^{2}}{c} \right)dk}.
  \label{eqn:LB_Fk_DefA}
\end{equation}
The value $P_{\ind{in}}$ specifies the mean power input by the spectral force. The total power input during one time step actually consists of two contributions
\begin{equation}
  P_{\ind{total}}=\frac{1}{2}\left< \vect{g}\left( t_{i} \right)\cdot \vect{g}\left( t_{i} \right) \right> + \left< \vect{g}\left( t_{i} \right)\cdot \vect{u}\left( t_{i} \right) \right> = P_{1}+P_{2}.
  \label{eqn:Ptotal}
\end{equation}
The first term is due to the force-force correlation and the second one is due to a force-velocity correlation. Since $\vect{u}$ is basically a random field, $P_{2}$ cannot be controlled and may become rather large. Therefore it is desirable to have $P_{2}\equiv 0$ at every time step and consequently $P_{1}=P_{\ind{in}}$. This can be achieved by demanding
\begin{equation}
  \hat{\vect{u}}\cdot\hat{\vect{g}}^{*}=0
  \label{eqn:udotg=0}
\end{equation}
for every active wave-vector. The condition \eqref{eqn:udotg=0} can be fulfilled if the angles $\theta_{1}$ and $\theta_{2}$ in Eq.~\eqref{eqn:LB_ArandBrand} are not chosen independently anymore. The angle $\theta_{1}$ is then given by
\begin{equation}
  \tan\theta_{1}=\frac{\sin\phi\Re\xi_{1}+\cos\phi\left( \sin\psi\Im\xi_{2}+\cos\psi\Re\xi_{2} \right)}{-\sin\phi\Im\xi_{1}+\cos\phi\left( \sin\psi\Re\xi_{2}-\cos\psi\Im\xi_{2} \right)},
  \label{eqn:tanTheta1}
\end{equation}
where $\xi_{1}=\hat{\vect{u}}\cdot\vect{e}_{1}$, $\xi_{2}=\hat{\vect{u}}\cdot\vect{e}_{2}$, and $\psi\in\left[ 0,2\pi \right]$ is a uniformly distributed random number. The angle $\theta_{2}$ is then calculated as $\theta_{2}=\psi-\theta_{1}$.

By specifying three input parameters we can determine all the properties of the spectral force $\vect{g}$:
\begin{itemize}
  \item The length $l_{\ind{g}}$, which defines the large scales of the turbulent field.
  \item The characteristic velocity $u_{\ind{g}}$, which fixes the timescale of the simulations.
  \item The Kolmogorov scale $\eta$ which determines the smallest turbulence scale.
\end{itemize}
To ensure that the simulated fluid field is incompressible, the Mach number has to be small. This gives the condition that the characteristic velocity has to be much smaller than the speed of sound, i.e.\ $u_{\ind{g}}\ll c_{\ind{s}}$.
From $l_{\ind{g}}$ we can determine $k_{\ind{f}}=n_{\ind{L}}/l_{\ind{g}}$, where $n_{\ind{L}}$ is the linear size of the LB lattice, and further $k_{\ind{a}}=1$ and $k_{\ind{b}}=2k_{\ind{f}}-1$. In Eq.~\eqref{eqn:LB_FkGaussian} the constant $c$ is chosen as $c=1$. The power input $P_{\ind{in}}$ is specified by $P_{\ind{in}}=u_{\ind{g}}^{3}/l_{\ind{g}}$ and the fluid viscosity is then given by
\begin{equation}
  \nu_{\ind{f}}=\left( P_{\ind{in}}\eta^{4} \right)^{1/3}=u_{\ind{g}}\left( \frac{\eta^{4}}{l_{\ind{g}}} \right)^{1/3}.
  \label{eqn:viscDef}
\end{equation}
From the viscosity the relaxation frequency in Eq.~\eqref{eqn:LBStandart} is determined as
\begin{equation}
  \nu_{\ind{f}}=c_{\ind{s}}^{2}\left( \frac{1}{\omega}-\frac{1}{2} \right) \quad\Leftrightarrow\quad \frac{1}{\omega}=\frac{\nu_{\ind{f}}}{c_{\ind{s}}^{2}}+\frac{1}{2}.
  \label{eqn:omegaFromVisc}
\end{equation}

At this point it is worth noting that there is no feedback from the particles to the fluid. According to Elghobashi \cite{Elghobashi1994} a four-way coupling would be needed in the range of Stokes numbers and volume fractions considered in this paper. Unfortunately incorporating a feedback force into the simulations is not trivial. Adding the negative drag force that acts on each of the particles to the spectral force $\vect{G}$ would be one way of coupling the fluid to the particles. The resulting forces would be singular and not necessarily located on a specific site of the LB lattice. Nash \textit{et al.}~\cite{Nash2008} used a regularized Dirac delta function to incorporate singular forces into the LB method in the case of low Reynolds numbers. The spectral force used in this paper is random and supplies energy to the whole system (on average) homogeneously distributed in space. The origin of the spectral force is completely artificial and energy conservation is therefore only fulfilled on average in time. Additionally, since the fluid velocity field is a random vector, the drag force and hence the feedback force are random vectors as well. Therefore including the feedback from the particles to the fluid would add an additional random component to the spectral force which could lead to an enhancement or a reduction of the local turbulent intensity. We assume that these effects cancel each other on average an thus the lack of the feedback force should not influence our results much.

\section{Simulation and results}
We used the method described in the last section with a lattice of linear size $n_{\ind{L}}=128$. The forcing length scale was $l_{\ind{g}}=32$, the characteristic fluid velocity $u_{\ind{g}}=0.05$ and the Kolmogorov length scale $\eta=0.5$. These values are the same as used by ten Cate \textit{et al.} \cite{Cate2006}. Using Eq.~\eqref{eqn:viscDef} this results in a fluid viscosity $\nu_{\ind{f}}=0.00625$. The fluid density was set to $\rho_{\ind{f}}=1.0$ and the initial fluid velocity was $\vect{u}=0$. After initialization we let the LB part of the simulation run until the mean kinetic energy of the fluid stayed constant over a certain time. Into this turbulent velocity field we put $n_{\ind{p}}$ particles of radius $R_{\ind{p}}=\eta$ on a regular grid. The initial velocities of the particles were chosen equal to the local fluid velocity. Since the particles can be located anywhere inside the system and are not confined to the lattice sites of the LB grid, the local fluid velocity was calculated by linear interpolation from the velocities of the eight surrounding lattice points. We simulated four different number of particles ($n_{\ind{p}}=884736$, $262144$, $110592$, and $32768$) corresponding to volume fractions of $\phi_{\ind{vol}}=22.1\%$, $6.5\%$, $2.8\%$, and $0.8\%$. Additionally the Stokes number of the particles was varied to be $\mathrm{St}=0.32$, $0.56$, $1.0$, $1.78$, $3.16$, $5.62$, $10.0$, $31.62$, and $100.0$. Using Eq.~\eqref{eqn:StokesNumber} it is possible to calculate the corresponding particle densities. The spring stiffness was set to $k_{\ind{n}}=25.0$ and the coefficient of restitution was varied as $e_{\ind{n}}=1.0$, $0.95$, $0.60$, and $0.25$. Additionally we performed simulations where particle collisions have been ignored and therefore particles could overlap and cross each other.
\begin{figure*}[!tbp]
  \includegraphics{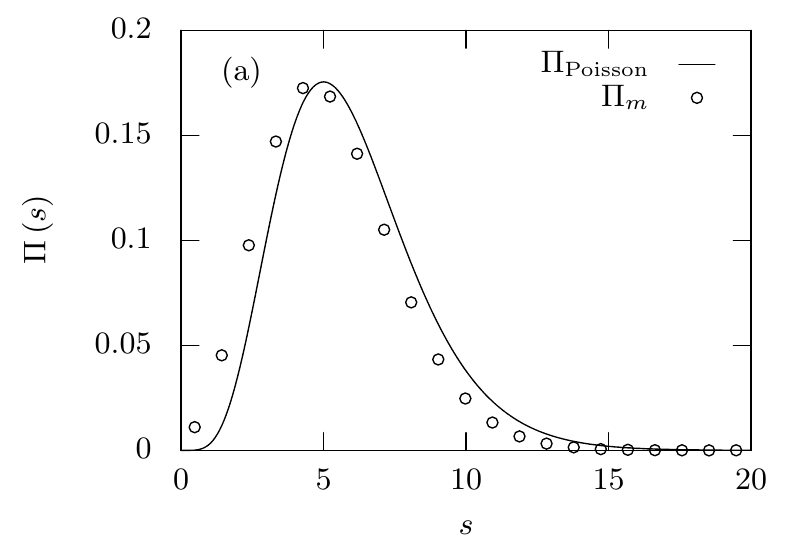}
  \hspace{0.5cm}
  \includegraphics{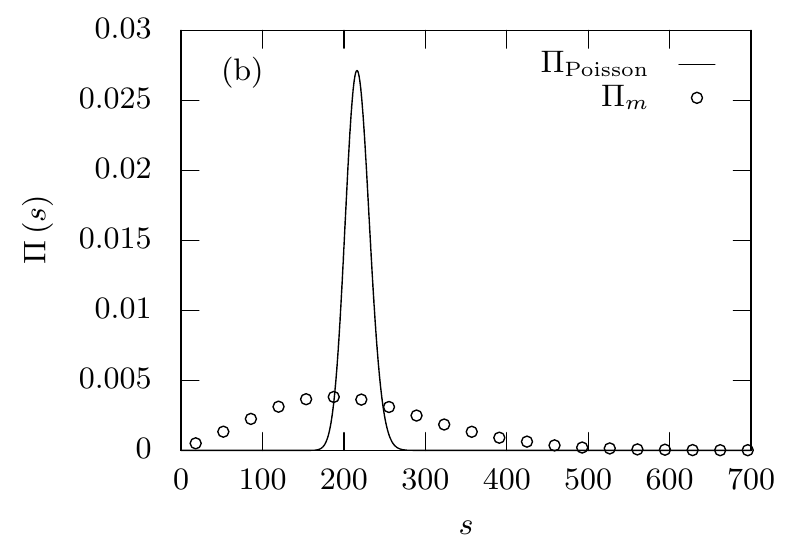}
  \caption{Histograms of the number of particles for different Stokes number and different box size. The left figure shows a case where the particle number distribution is close to a Poisson distribution which means, that particles are uniformly distributed in space. On the other hand the right figure shows a case where the distribution is much broader and differs strongly from a Poisson distribution. In this case preferential concentration occurs.}
  \label{fig:hist}
\end{figure*}
To quantify the clustering of particles we follow the work of Fessler \textit{et al.} \cite{Fessler1994}. The main idea of this analysis is to measure the local deviation of the particle number from a Poisson distribution. This is reasonable since if one divides a cubic volume where particles are uniformly distributed in space in smaller cubes, the number of particles in these smaller boxes are Poisson distributed. In more detail we take a snapshot of the system, divide it into $m^{3}$ boxes of linear size $l_{m}=n_{\ind{L}}/m$, ($m=4,\dots ,n_{\ind{L}}$) and then count the number of particles $s$ in each box. For every box size we then measure the mean $\mu_{m}$, the standard deviation $\sigma_{m}$, as well as the distribution $\Pi_{m}\left( s \right)$ of the number of particles. If the particles are uniformly distributed in space, $\Pi_{m}\left( s \right)$ should be a Poisson distribution
\begin{equation}
  \Pi_{\ind{Poisson}}\left( s;\lambda = \mu_{m} \right) = \frac{\lambda^{s}e^{-\lambda}}{s!}.
  \label{eqn:PoissonDistributions}
\end{equation}
Fig.~\ref{fig:hist} shows two examples of these distributions. In Fig.~\ref{fig:hist}~(a) the measured distribution is close to the Poisson distribution and therefore the particles are almost uniformly distributed in space, i.e. no preferential concentration is visible. In Fig.~\ref{fig:hist}~(b) on the other hand the measured distribution is much broader than a Poisson distribution with the same mean value $\mu_{m}$. This means that there are boxes with too large or too small numbers of particles to be compatible with a Poisson distribution. Therefore particles are clustered in small regions of space, i.e. preferential concentration occurs.
\begin{figure*}[!tbp]
  \includegraphics{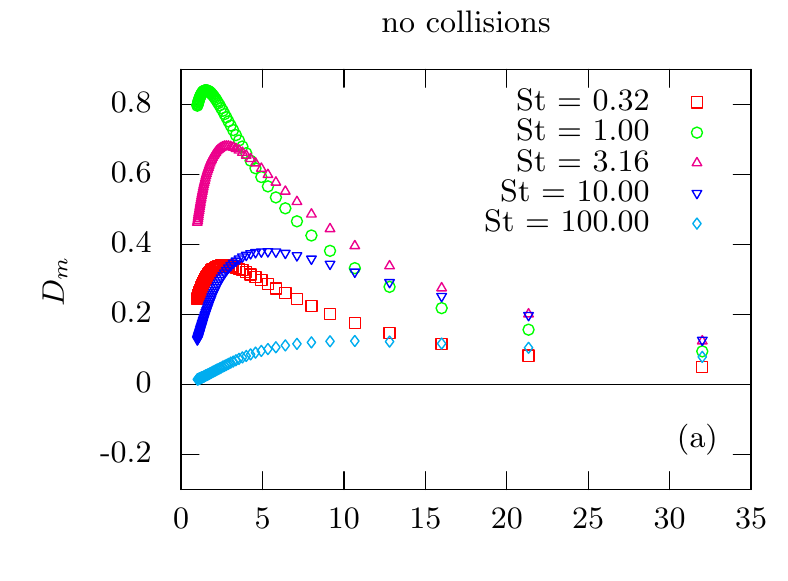}\hspace{0.5cm}
  \includegraphics{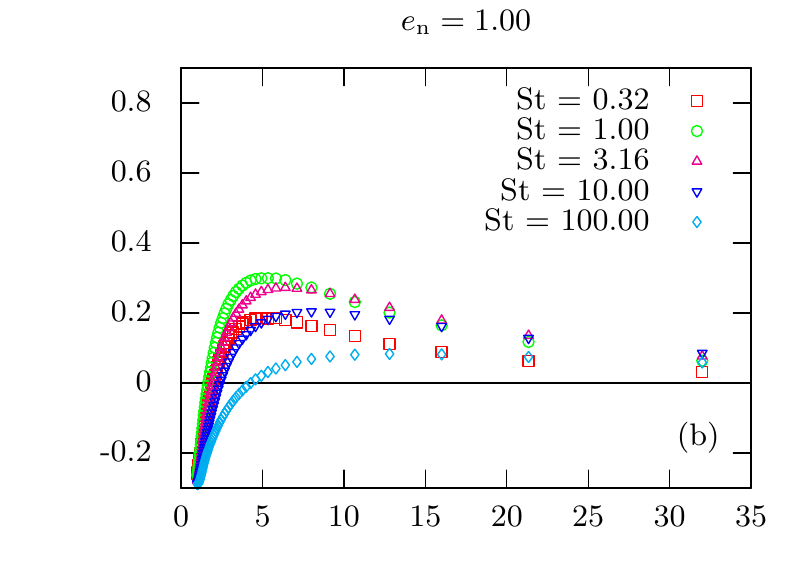}\\[0.2cm]
  \includegraphics{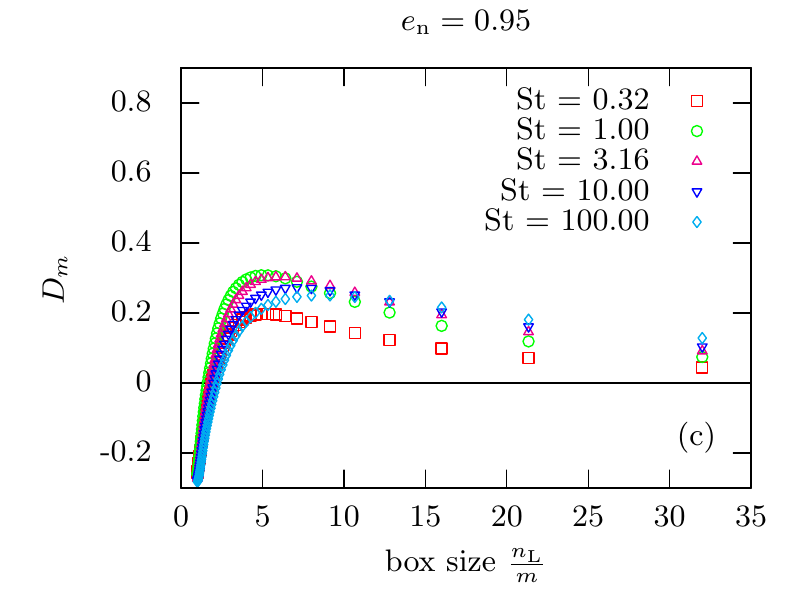}\hspace{0.5cm}
  \includegraphics{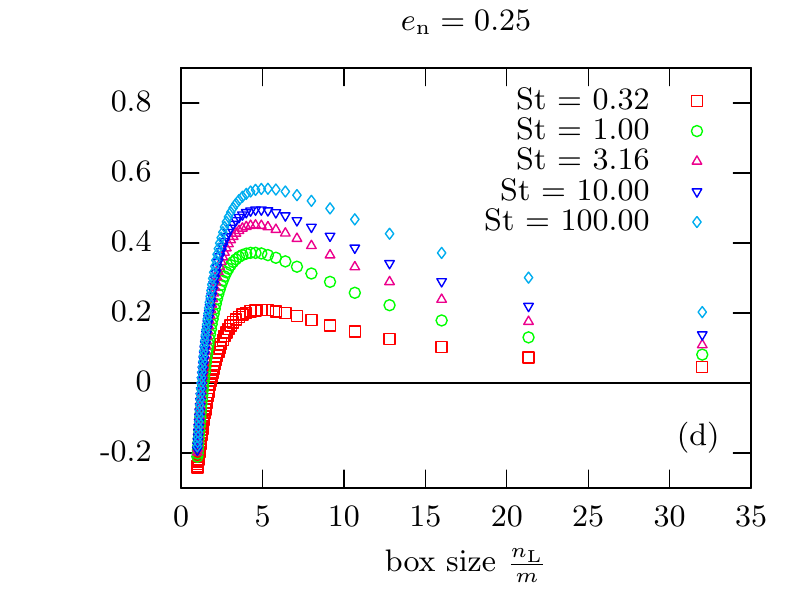}
  \caption{Dependence of $D_{m}$ on the box size $\frac{n_{\ind{L}}}{m}$ for different Stokes numbers $\mathrm{St}$ with and without collisions at a volume fraction $\phi_{\ind{vol}}=22.1\%$. Particles are randomly distributed if $D_{m}=0$. If $D_{m}>0$ preferential concentration occurs and if $D_{m}<0$, on average, the same number of particles is located in boxes of corresponding size. (a) Simulations \emph{without} collisions. Preferential concentration is maximal for $\mathrm{St}\approx 1$. (b) Simulations \emph{with} collisions and coefficient of restitution $e_{\ind{n}}=1.0$, which corresponds to elastic collisions. Preferential concentration is reduced but the maximum around $\mathrm{St}\approx 1$ is still visible. (c) Simulations with $e_{\ind{n}}=0.95$. Clustering of particles is less dependent on the Stokes number, but the maximum of preferential concentration is still visible. (d) Simulations for $e_{\ind{n}}=0.25$. Dissipative collisions dominate the clustering of particles. Since particles with higher Stokes number are less influenced by the turbulent fluid, clustering dominates.}
  \label{fig:D}
\end{figure*}
To further quantify the deviation from a Poisson distribution it is useful to define \cite{Fessler1994} the value
\begin{equation}
  D_{m}=\frac{\sigma_{m} - \sigma_{\ind{Poisson}}}{\mu_{m}}.
  \label{eq:D_Def}
\end{equation}
Depending on the value of $D_{m}$ one can distinguish three regimes:
\begin{description}
  \item[$D_{m}>0$] The distribution $\Pi_{m}\left( s \right)$ is broader than $\Pi_{\ind{Poisson}}\left( s;\mu_{m} \right)$ and therefore clustering occurs. The larger $D_{m}$ is the more pronounced this effect is.
  \item[$D_{m}=0$] The distribution $\Pi_{m}\left( s \right)$ and $\Pi_{\ind{Poisson}}\left( s;\mu_{m} \right)$ are equally broad. Particles are uniformly distributed in space.
  \item[$D_{m}<0$] The distribution $\Pi_{m}\left( s \right)$ is narrower than a corresponding Poisson distribution. This means that all boxes contain more or less the same number of particles.
\end{description}

We first applied this analysis to the case where collisions between particles were ignored. The behavior of $D_{m}$ for particles with different Stokes numbers $\mathrm{St}$ at a volume fraction of $\phi_{\ind{vol}}=22.1\%$ is shown in Fig.~\ref{fig:D}~(a). The deviation of the particle distribution from a Poisson distribution is clearly visible and as expected \cite{Fessler1994} preferential concentration is strongest at a Stokes numbers around $\mathrm{St}\approx 1$.

In a next step we included collisions in the simulations. We first set the coefficient of restitution to $e_{\ind{n}}=1.0$, which corresponds to elastic collisions. The results of this case are shown in Fig.~\ref{fig:D}~(b). The collisions reduce the strength of the preferential concentration, but the strongest clustering is still visible for a Stokes number $\mathrm{St}\approx 1$. This effect is to be expected since collisions introduce a mechanism that tries to move particles apart and counteracts the ``attraction'' of particles in regions of low vorticity and high strain. Therefore the overall strength of preferential concentration is reduced. Additionally $D_{m}$ even becomes negative for small box sizes $n_{\ind{L}}/m$. In contrast to the collisonless case, particles repel each other and cannot come arbitrarily close together. For box sizes close to the particle diameter this means that fluctuations in the particle number are strongly reduced. Therefore $D_{m}$ can become negative for these box sizes.

Now we reduced the coefficient of restitution. As shown in Refs.~\cite{Luding1999,Miller2004} the influence of the dissipation during particle collisions should already influence the clustering at moderate values of $e_{\ind{n}}$. Therefore we chose $e_{\ind{n}}=0.95$. The results of these simulations are shown in Fig.~\ref{fig:D}~(c). One can see that the dependence of the clustering on the Stokes number is less pronounced in this case, since the curves of $D_{m}$ are rather close to each other for different Stokes numbers. Even so maximal preferential concentration at $\mathrm{St}\approx 1$ can still be observed.

Fig.~\ref{fig:D}~(d) finally shows the results of simulations where the coefficient of restitution was set to $e_{\ind{n}}=0.25$. Here the maximal preferential concentration at $\mathrm{St}\approx 1$ cannot be observed anymore, and the clustering is clearly dominated by particle collisions. The effect of clustering increases with increasing Stokes number, since particles with higher $\mathrm{St}$ are less influenced by the fluid velocity field.

\begin{figure}[!tbp]
  \includegraphics{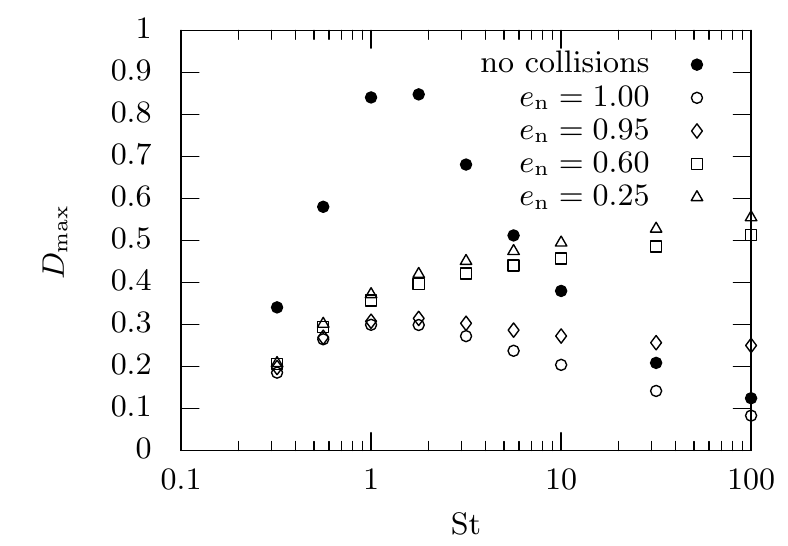}
  \caption{Plot of the maximum of $D_{m}$ for different Stokes numbers $\mathrm{St}$ at a volume fraction of $\phi_{\ind{vol}}=22.1\%$. The simulation without collisions clearly shows the strongest preferential concentration at $\mathrm{St}\approx 1$. This effect is reduced by collisions. For a coefficient of restitution of $e_{\ind{n}}=0.95$ a maximum at $\mathrm{St}\approx 1$ is still visible. For smaller $e_{\ind{n}}$ this maximum disappears and the clustering becomes larger for larger Stokes number. The reason for this is that the fluid velocity field less and less influences the motion of the particles and the collisions become dominant.}
  \label{fig:Dmax}
\end{figure}

To further investigate the two regimes of preferential concentration due to turbulence and clustering due to collisional cooling we again followed Ref.~\cite{Fessler1994} and determined the maximum of $D_{m}$. In Fig.~\ref{fig:Dmax} we then plot the values of $D_{\ind{max}}$ versus $\mathrm{St}$ for different coefficients of restitution and also the collisionless case. In the latter case preferential concentration is most pronounced. This is clear since in this case there is no effect which tries to move particles apart. This means particle can overlap and therefore many particles can concentrate in regions of the fluid velocity field where vorticity is low and strain is high. When particle collisions are included in the simulation, large overlaps between particles are not allowed anymore. Therefore preferential concentration is reduced. This effect can also be seen in Fig.~\ref{fig:Dmax} for $e_{\ind{n}}=1.0$. The maximum at $\mathrm{St}\approx 1$ is still clearly visible but the overall strength of preferential concentration is reduced. When the coefficient of restitution is reduced, another effect for the clustering of particles is introduced. Therefore $D_{\ind{max}}$ should increase when $e_{\ind{n}}$ is reduced. This effect can be seen in Fig.~\ref{fig:Dmax} for $e_{\ind{n}}=0.95$. The increase of $D_{\ind{max}}$ is larger for higher Stokes numbers. This can be explained by the fact, that particles with larger $\mathrm{St}$ are less influenced by the fluid velocity field. Further decreasing $e_{\ind{n}}$ brings us into a regime where the dissipative particle collisions are the dominant mechanism for clustering of particles. Fig.~\ref{fig:Dmax} shows the behavior of $D_{\ind{max}}$ for $e_{\ind{n}}=0.6$ and $e_{\ind{n}}=0.25$.

\begin{figure}[!tbp]
  \includegraphics{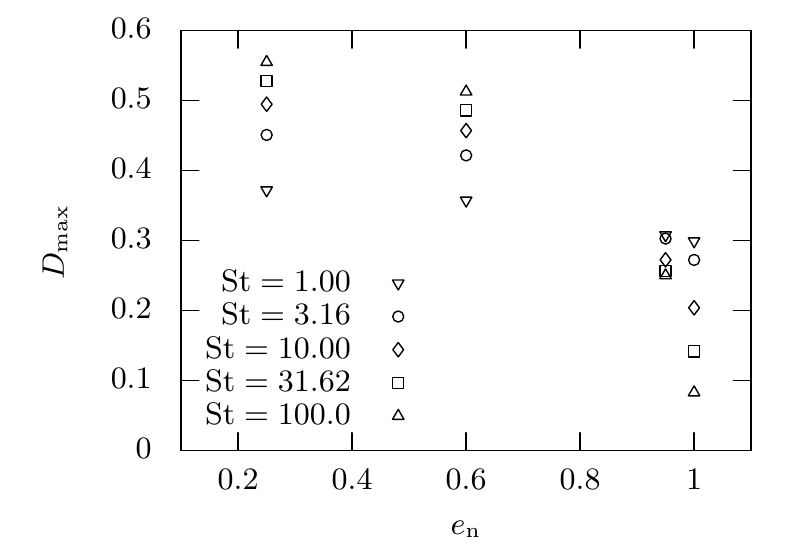}
  \caption{The dependence of $D_{\ind{max}}$ on $e_{\ind{n}}$ for different Stokes numbers. (Same data as in Fig.~\ref{fig:Dmax}) At $e_{\ind{n}}\lessapprox 0.95$ a crossover is visible. For $e_{\ind{n}}$ above the crossover the influence of the turbulent velocity field is still visible and preferential concentration occurs; below, the main mechanism for clustering are the dissipative particle collisions.}
  \label{fig:DmaxAtSt}
\end{figure}

Fig.~\ref{fig:DmaxAtSt} shows the same data as Fig.~\ref{fig:Dmax} but this time exhibiting the dependence of $D_{\ind{max}}$ on $e_{\ind{n}}$ for different Stokes numbers. The plot shows a crossover at a coefficient of restitution around $e_{\ind{n}}\lessapprox 0.95$. Above this threshold the clustering is influenced by the turbulent velocity field and preferential concentration at $\mathrm{St}\approx 1$ can be observed. Below the threshold the collisions between particles are the dominant mechanism for clustering of particles.

\begin{figure*}[!tbp]
  \includegraphics{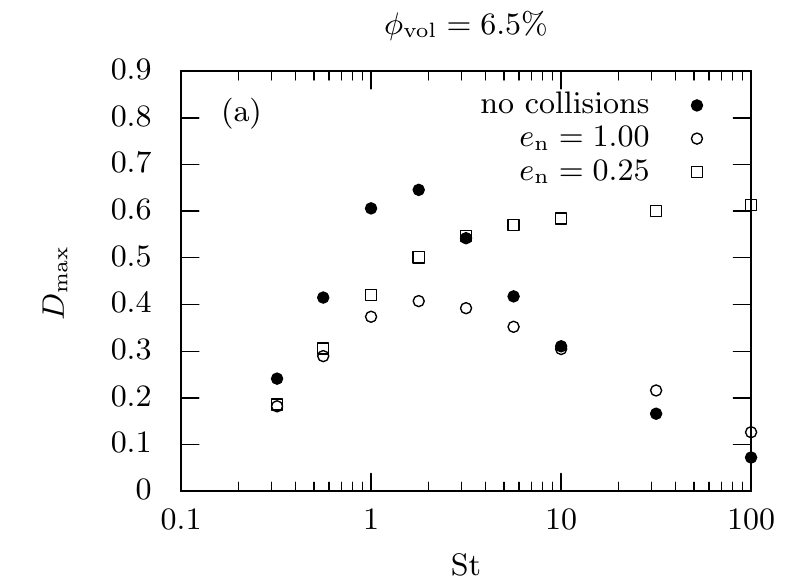}\hspace{0.5cm}
  \includegraphics{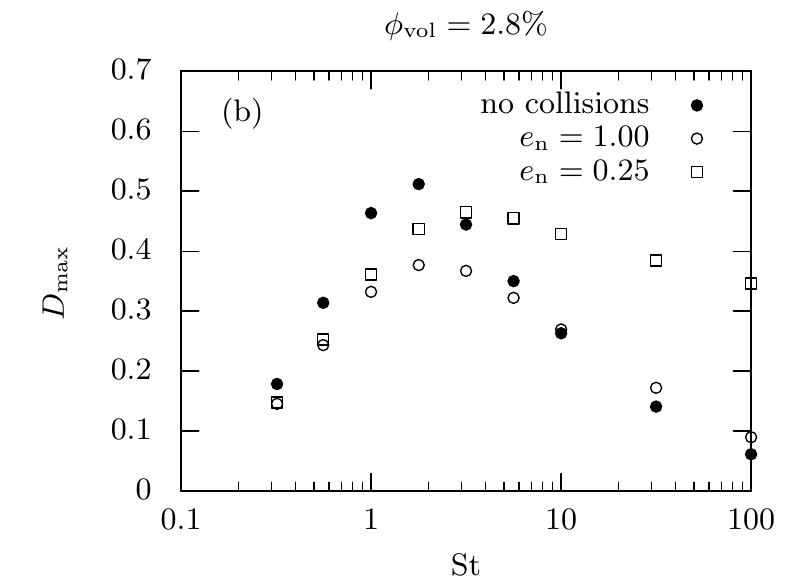}\\
  \includegraphics{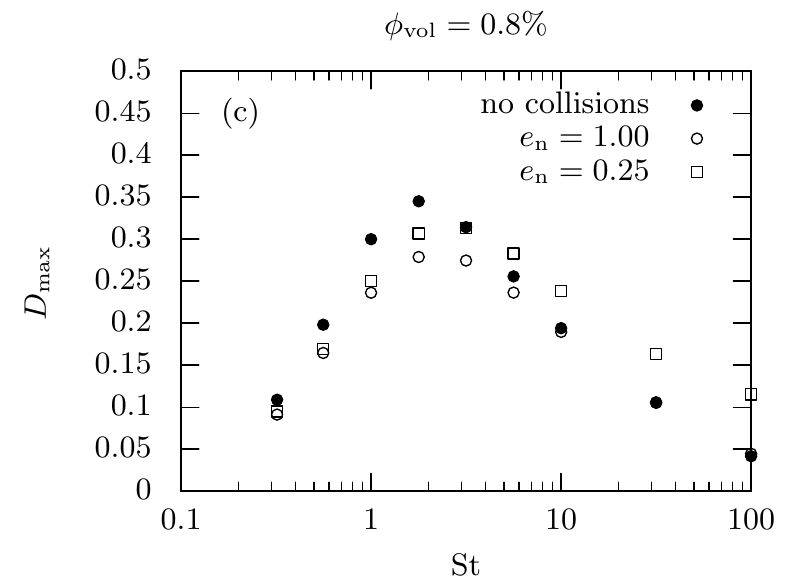}
  \caption{Dependence of $D_{\ind{max}}$ on the Stokes number $\mathrm{St}$ for different coefficients of restitution $e_{\ind{n}}$ and the collisionless case at different volume fractions $\phi_{\ind{vol}}$. Even at moderate volume fraction the influence of the dissipative particle collisions is clearly visible. Only for densities below $1\%$ the effect becomes smaller. Also visible in this figure is the difference between the results of the simulations with elastic collisions ($e_{\ind{n}}=1.0$) and the collisionless case.}
  \label{fig:DmaxDensity}
\end{figure*}
We further investigated the influence of the particle volume fraction $\phi_{\ind{vol}}$ on the clustering of particles. By reducing the number of particles $n_{\ind{p}}$ we simulated systems with volume fractions $\phi_{\ind{vol}}=6.4\%$, $2.8\%$, and $0.8\%$. Again we varied the Stokes number of the particles and measured $D_{m}$ for different box sizes $\frac{n_{\ind{L}}}{m}$. We then determined the maximum of $D_{m}$ and plotted the results in Fig.~\ref{fig:DmaxDensity}. The first thing to notice in these plots is the influence of the particle collisions. As expected the difference between the result of the simulations without collisions and the case of elastic collisions ($e_{\ind{n}}=1.0$) becomes smaller with lower volume fractions. The difference between these two cases is less distinct for particles with larger Stokes numbers and more pronounced for $\mathrm{St}\approx 1$. For $\phi_{\ind{vol}}=0.8\%$, the difference between these two cases even disappears for Stokes number above $10$.

The next point to mention is that even at moderate volume fractions the influence of the dissipative particle collisions is still visible. At $\phi_{\ind{vol}}=6.5\%$ with $e_{\ind{n}}=0.25$ the clustering is still dominated by the particle collisions. At the lower volume fraction $\phi_{\ind{vol}}=2.8\%$ the influence of the turbulent velocity field can be seen by a maximum of $D_{\ind{max}}$ at $\mathrm{St}\approx 1$, but the effect of the dissipative particle collisions is recognizable at larger Stokes numbers. Only at rather small volume fraction of $\phi_{\ind{vol}}=0.8\%$ the influence of the particle collisions becomes negligible.

As already explained in Section~\ref{sec:model} the model does not implement any feedback from the particles onto the fluid. The influence of the particles on the turbulent velocity field is known in the literature as turbulence modulation~\cite{Balachandar2010}. Depending on several parameters like particle radius, Stokes number, volume fraction, Reynolds number, etc., this modification may be either an enhancement or a reduction of the turbulence intensity, which in turn can influence the local structure of preferential concentration. Despite several experimental, numerical and theoretical investigations a conclusive understanding of the involved phenomena is still not found~\cite{Balachandar2010}. It is reasonable to assume that the modification is less important for lower volume fractions. Apart from that, it is rather difficult to reliably predict the changes due to turbulence modulation. Recent high-resolution simulations~\cite{Burton2005} and highly resolved particle image velocimetry experiments~\cite{Tanaka2010} show that for particles with diameter $d\gtrapprox \eta$ dissipation around the particles is strongly enhanced which then leads to a reduction in turbulence intensity in this regions. Including a feedback force in our simulations may therefore lead to a reduction of preferential concentration.

\section{Conclusion}
In this paper we used a DEM together with a LB method to simulate the motion of inelastically colliding soft spheres in a homogeneous turbulent flow field. We investigated the influence of dissipative particle collisions on the clustering of particles and found that already at low densities collisions can be an important factor. For volume fractions around $22.1\%$ collisions become dominant below a coefficient of restitution of $e_{\ind{n}}\lessapprox 0.95$. Above this threshold we observed preferential concentration with a maximum for particles with a Stokes number around $\mathrm{St}\approx 1$. For lower volume fractions the influence of particle collisions becomes less important and the crossover between preferential concentration and collisional cooling is shifted to smaller values of $e_{\ind{n}}$.

\begin{acknowledgments}
  The authors thank M. Mendoza and S. Succi for their useful hints and interesting discussions. We also acknowledge financial support from ETH Research Grant \mbox{ETH-06 11-1}.
\end{acknowledgments}

\bibliography{references}

\begin{thebibliography}{30}%
\makeatletter
\providecommand \@ifxundefined [1]{%
 \@ifx{#1\undefined}
}%
\providecommand \@ifnum [1]{%
 \ifnum #1\expandafter \@firstoftwo
 \else \expandafter \@secondoftwo
 \fi
}%
\providecommand \@ifx [1]{%
 \ifx #1\expandafter \@firstoftwo
 \else \expandafter \@secondoftwo
 \fi
}%
\providecommand \natexlab [1]{#1}%
\providecommand \enquote  [1]{``#1''}%
\providecommand \bibnamefont  [1]{#1}%
\providecommand \bibfnamefont [1]{#1}%
\providecommand \citenamefont [1]{#1}%
\providecommand \href@noop [0]{\@secondoftwo}%
\providecommand \href [0]{\begingroup \@sanitize@url \@href}%
\providecommand \@href[1]{\@@startlink{#1}\@@href}%
\providecommand \@@href[1]{\endgroup#1\@@endlink}%
\providecommand \@sanitize@url [0]{\catcode `\\12\catcode `\$12\catcode
  `\&12\catcode `\#12\catcode `\^12\catcode `\_12\catcode `\%12\relax}%
\providecommand \@@startlink[1]{}%
\providecommand \@@endlink[0]{}%
\providecommand \url  [0]{\begingroup\@sanitize@url \@url }%
\providecommand \@url [1]{\endgroup\@href {#1}{\urlprefix }}%
\providecommand \urlprefix  [0]{URL }%
\providecommand \Eprint [0]{\href }%
\providecommand \doibase [0]{http://dx.doi.org/}%
\providecommand \selectlanguage [0]{\@gobble}%
\providecommand \bibinfo  [0]{\@secondoftwo}%
\providecommand \bibfield  [0]{\@secondoftwo}%
\providecommand \translation [1]{[#1]}%
\providecommand \BibitemOpen [0]{}%
\providecommand \bibitemStop [0]{}%
\providecommand \bibitemNoStop [0]{.\EOS\space}%
\providecommand \EOS [0]{\spacefactor3000\relax}%
\providecommand \BibitemShut  [1]{\csname bibitem#1\endcsname}%
\let\auto@bib@innerbib\@empty
\bibitem [{\citenamefont {Eaton}\ and\ \citenamefont
  {Fessler}(1994)}]{Eaton1994}%
  \BibitemOpen
  \bibfield  {author} {\bibinfo {author} {\bibfnamefont {J.~K.}\ \bibnamefont
  {Eaton}}\ and\ \bibinfo {author} {\bibfnamefont {J.~R.}\ \bibnamefont
  {Fessler}},\ }\href {\doibase 10.1016/0301-9322(94)90072-8} {\bibfield
  {journal} {\bibinfo  {journal} {International Journal of Multiphase Flow}\
  }\textbf {\bibinfo {volume} {20}},\ \bibinfo {pages} {169} (\bibinfo {year}
  {1994})}\BibitemShut {NoStop}%
\bibitem [{\citenamefont {Fessler}\ \emph {et~al.}(1994)\citenamefont
  {Fessler}, \citenamefont {Kulick},\ and\ \citenamefont
  {Eaton}}]{Fessler1994}%
  \BibitemOpen
  \bibfield  {author} {\bibinfo {author} {\bibfnamefont {J.~R.}\ \bibnamefont
  {Fessler}}, \bibinfo {author} {\bibfnamefont {J.~D.}\ \bibnamefont {Kulick}},
  \ and\ \bibinfo {author} {\bibfnamefont {J.~K.}\ \bibnamefont {Eaton}},\
  }\href {\doibase 10.1063/1.868445} {\bibfield  {journal} {\bibinfo  {journal}
  {Physics of Fluids}\ }\textbf {\bibinfo {volume} {6}},\ \bibinfo {pages}
  {3742} (\bibinfo {year} {1994})}\BibitemShut {NoStop}%
\bibitem [{\citenamefont {Bec}\ \emph {et~al.}(2007)\citenamefont {Bec},
  \citenamefont {Biferale}, \citenamefont {Cencini}, \citenamefont {Lanotte},
  \citenamefont {Musacchio},\ and\ \citenamefont {Toschi}}]{Bec2007}%
  \BibitemOpen
  \bibfield  {author} {\bibinfo {author} {\bibfnamefont {J.}~\bibnamefont
  {Bec}}, \bibinfo {author} {\bibfnamefont {L.}~\bibnamefont {Biferale}},
  \bibinfo {author} {\bibfnamefont {M.}~\bibnamefont {Cencini}}, \bibinfo
  {author} {\bibfnamefont {A.}~\bibnamefont {Lanotte}}, \bibinfo {author}
  {\bibfnamefont {S.}~\bibnamefont {Musacchio}}, \ and\ \bibinfo {author}
  {\bibfnamefont {F.}~\bibnamefont {Toschi}},\ }\href {\doibase
  10.1103/PhysRevLett.98.084502} {\bibfield  {journal} {\bibinfo  {journal}
  {Physical Review Letters}\ }\textbf {\bibinfo {volume} {98}},\ \bibinfo
  {pages} {084502} (\bibinfo {year} {2007})}\BibitemShut {NoStop}%
\bibitem [{\citenamefont {Balachandar}\ and\ \citenamefont
  {Eaton}(2010)}]{Balachandar2010}%
  \BibitemOpen
  \bibfield  {author} {\bibinfo {author} {\bibfnamefont {S.}~\bibnamefont
  {Balachandar}}\ and\ \bibinfo {author} {\bibfnamefont {J.~K.}\ \bibnamefont
  {Eaton}},\ }\href {\doibase 10.1146/annurev.fluid.010908.165243} {\bibfield
  {journal} {\bibinfo  {journal} {Annual Review of Fluid Mechanics}\ }\textbf
  {\bibinfo {volume} {42}},\ \bibinfo {pages} {111} (\bibinfo {year}
  {2010})}\BibitemShut {NoStop}%
\bibitem [{\citenamefont {Collins}\ and\ \citenamefont
  {Keswani}(2004)}]{Collins2004}%
  \BibitemOpen
  \bibfield  {author} {\bibinfo {author} {\bibfnamefont {L.~R.}\ \bibnamefont
  {Collins}}\ and\ \bibinfo {author} {\bibfnamefont {A.}~\bibnamefont
  {Keswani}},\ }\href {\doibase 10.1088/1367-2630/6/1/119} {\bibfield
  {journal} {\bibinfo  {journal} {New Journal of Physics}\ }\textbf {\bibinfo
  {volume} {6}},\ \bibinfo {pages} {119} (\bibinfo {year} {2004})}\BibitemShut
  {NoStop}%
\bibitem [{\citenamefont {Cencini}\ \emph {et~al.}(2006)\citenamefont
  {Cencini}, \citenamefont {Bec}, \citenamefont {Biferale}, \citenamefont
  {Boffetta}, \citenamefont {Lanotte}, \citenamefont {Musacchio},\ and\
  \citenamefont {Toschi}}]{Cencini2006}%
  \BibitemOpen
  \bibfield  {author} {\bibinfo {author} {\bibfnamefont {M.}~\bibnamefont
  {Cencini}}, \bibinfo {author} {\bibfnamefont {J.}~\bibnamefont {Bec}},
  \bibinfo {author} {\bibfnamefont {L.}~\bibnamefont {Biferale}}, \bibinfo
  {author} {\bibfnamefont {G.}~\bibnamefont {Boffetta}}, \bibinfo {author}
  {\bibfnamefont {A.~S.}\ \bibnamefont {Lanotte}}, \bibinfo {author}
  {\bibfnamefont {S.}~\bibnamefont {Musacchio}}, \ and\ \bibinfo {author}
  {\bibfnamefont {F.}~\bibnamefont {Toschi}},\ }\href {\doibase
  10.1080/14685240600675727} {\bibfield  {journal} {\bibinfo  {journal}
  {Journal of Turbulence}\ }\textbf {\bibinfo {volume} {7}},\ \bibinfo {pages}
  {36} (\bibinfo {year} {2006})}\BibitemShut {NoStop}%
\bibitem [{\citenamefont {Bec}(2005)}]{Bec2005}%
  \BibitemOpen
  \bibfield  {author} {\bibinfo {author} {\bibfnamefont {J.}~\bibnamefont
  {Bec}},\ }\href {\doibase 10.1017/S0022112005003368} {\bibfield  {journal}
  {\bibinfo  {journal} {Journal of Fluid Mechanics}\ }\textbf {\bibinfo
  {volume} {528}},\ \bibinfo {pages} {255} (\bibinfo {year}
  {2005})}\BibitemShut {NoStop}%
\bibitem [{\citenamefont {Bec}\ \emph {et~al.}(2006)\citenamefont {Bec},
  \citenamefont {Biferale}, \citenamefont {Boffetta}, \citenamefont {Celani},
  \citenamefont {Cencini}, \citenamefont {Lanotte}, \citenamefont {Musacchio},\
  and\ \citenamefont {Toschi}}]{Bec2006}%
  \BibitemOpen
  \bibfield  {author} {\bibinfo {author} {\bibfnamefont {J.}~\bibnamefont
  {Bec}}, \bibinfo {author} {\bibfnamefont {L.}~\bibnamefont {Biferale}},
  \bibinfo {author} {\bibfnamefont {G.}~\bibnamefont {Boffetta}}, \bibinfo
  {author} {\bibfnamefont {A.}~\bibnamefont {Celani}}, \bibinfo {author}
  {\bibfnamefont {M.}~\bibnamefont {Cencini}}, \bibinfo {author} {\bibfnamefont
  {A.}~\bibnamefont {Lanotte}}, \bibinfo {author} {\bibfnamefont
  {S.}~\bibnamefont {Musacchio}}, \ and\ \bibinfo {author} {\bibfnamefont
  {F.}~\bibnamefont {Toschi}},\ }\href {\doibase 10.1017/S002211200500844X}
  {\bibfield  {journal} {\bibinfo  {journal} {Journal of Fluid Mechanics}\
  }\textbf {\bibinfo {volume} {550}},\ \bibinfo {pages} {349} (\bibinfo {year}
  {2006})}\BibitemShut {NoStop}%
\bibitem [{\citenamefont {Biferale}\ \emph {et~al.}(2006)\citenamefont
  {Biferale}, \citenamefont {Boffetta}, \citenamefont {Celani}, \citenamefont
  {Lanotte},\ and\ \citenamefont {Toschi}}]{Biferale2006}%
  \BibitemOpen
  \bibfield  {author} {\bibinfo {author} {\bibfnamefont {L.}~\bibnamefont
  {Biferale}}, \bibinfo {author} {\bibfnamefont {G.}~\bibnamefont {Boffetta}},
  \bibinfo {author} {\bibfnamefont {A.}~\bibnamefont {Celani}}, \bibinfo
  {author} {\bibfnamefont {A.}~\bibnamefont {Lanotte}}, \ and\ \bibinfo
  {author} {\bibfnamefont {F.}~\bibnamefont {Toschi}},\ }\href {\doibase
  10.1080/14685240500460832} {\bibfield  {journal} {\bibinfo  {journal}
  {Journal of Turbulence}\ }\textbf {\bibinfo {volume} {7}},\ \bibinfo {pages}
  {N6} (\bibinfo {year} {2006})}\BibitemShut {NoStop}%
\bibitem [{\citenamefont {Luding}\ and\ \citenamefont
  {Herrmann}(1999)}]{Luding1999}%
  \BibitemOpen
  \bibfield  {author} {\bibinfo {author} {\bibfnamefont {S.}~\bibnamefont
  {Luding}}\ and\ \bibinfo {author} {\bibfnamefont {H.~J.}\ \bibnamefont
  {Herrmann}},\ }\href {\doibase 10.1063/1.166441} {\bibfield  {journal}
  {\bibinfo  {journal} {Chaos}\ }\textbf {\bibinfo {volume} {9}},\ \bibinfo
  {pages} {673} (\bibinfo {year} {1999})}\BibitemShut {NoStop}%
\bibitem [{\citenamefont {Miller}\ and\ \citenamefont
  {Luding}(2004)}]{Miller2004}%
  \BibitemOpen
  \bibfield  {author} {\bibinfo {author} {\bibfnamefont {S.}~\bibnamefont
  {Miller}}\ and\ \bibinfo {author} {\bibfnamefont {S.}~\bibnamefont
  {Luding}},\ }\href {\doibase 10.1103/PhysRevE.69.031305} {\bibfield
  {journal} {\bibinfo  {journal} {Physical Review E}\ }\textbf {\bibinfo
  {volume} {69}},\ \bibinfo {pages} {031305} (\bibinfo {year}
  {2004})}\BibitemShut {NoStop}%
\bibitem [{\citenamefont {Alvelius}(1999)}]{Alvelius1999}%
  \BibitemOpen
  \bibfield  {author} {\bibinfo {author} {\bibfnamefont {K.}~\bibnamefont
  {Alvelius}},\ }\href {\doibase 10.1063/1.870050} {\bibfield  {journal}
  {\bibinfo  {journal} {Physics of Fluids}\ }\textbf {\bibinfo {volume} {11}},\
  \bibinfo {pages} {1880} (\bibinfo {year} {1999})}\BibitemShut {NoStop}%
\bibitem [{\citenamefont {ten Cate}\ \emph {et~al.}(2006)\citenamefont {ten
  Cate}, \citenamefont {van Vliet}, \citenamefont {Derksen},\ and\
  \citenamefont {{Van den Akker}}}]{Cate2006}%
  \BibitemOpen
  \bibfield  {author} {\bibinfo {author} {\bibfnamefont {A.}~\bibnamefont {ten
  Cate}}, \bibinfo {author} {\bibfnamefont {E.}~\bibnamefont {van Vliet}},
  \bibinfo {author} {\bibfnamefont {J.~J.}\ \bibnamefont {Derksen}}, \ and\
  \bibinfo {author} {\bibfnamefont {H.~E.~A.}\ \bibnamefont {{Van den
  Akker}}},\ }\href {\doibase 10.1016/j.compfluid.2005.06.001} {\bibfield
  {journal} {\bibinfo  {journal} {Computers \& Fluids}\ }\textbf {\bibinfo
  {volume} {35}},\ \bibinfo {pages} {1239} (\bibinfo {year}
  {2006})}\BibitemShut {NoStop}%
\bibitem [{\citenamefont {Cundall}\ and\ \citenamefont
  {Strack}(1979)}]{Cundall1979}%
  \BibitemOpen
  \bibfield  {author} {\bibinfo {author} {\bibfnamefont {P.~A.}\ \bibnamefont
  {Cundall}}\ and\ \bibinfo {author} {\bibfnamefont {O.~D.~L.}\ \bibnamefont
  {Strack}},\ }\href {\doibase 10.1680/geot.1979.29.1.47} {\bibfield  {journal}
  {\bibinfo  {journal} {G\'{e}otechnique}\ }\textbf {\bibinfo {volume} {29}},\
  \bibinfo {pages} {47} (\bibinfo {year} {1979})}\BibitemShut {NoStop}%
\bibitem [{\citenamefont {Herrmann}\ and\ \citenamefont
  {Luding}(1998)}]{Herrmann1998}%
  \BibitemOpen
  \bibfield  {author} {\bibinfo {author} {\bibfnamefont {H.~J.}\ \bibnamefont
  {Herrmann}}\ and\ \bibinfo {author} {\bibfnamefont {S.}~\bibnamefont
  {Luding}},\ }\href {\doibase 10.1007/s001610050089} {\bibfield  {journal}
  {\bibinfo  {journal} {Continuum Mechanics and Thermodynamics}\ }\textbf
  {\bibinfo {volume} {10}},\ \bibinfo {pages} {189} (\bibinfo {year}
  {1998})}\BibitemShut {NoStop}%
\bibitem [{\citenamefont {Luding}(1998)}]{Luding1998}%
  \BibitemOpen
  \bibfield  {author} {\bibinfo {author} {\bibfnamefont {S.}~\bibnamefont
  {Luding}},\ }in\ \href@noop {} {\emph {\bibinfo {booktitle} {Physics of dry
  granular Media}}},\ \bibinfo {editor} {edited by\ \bibinfo {editor}
  {\bibfnamefont {H.~J.}\ \bibnamefont {Herrmann}}, \bibinfo {editor}
  {\bibfnamefont {J.-P.}\ \bibnamefont {Hovi}}, \ and\ \bibinfo {editor}
  {\bibfnamefont {S.}~\bibnamefont {Luding}}}\ (\bibinfo  {publisher} {Kluwer
  Academic Publishers},\ \bibinfo {address} {Dordrecht},\ \bibinfo {year}
  {1998})\BibitemShut {NoStop}%
\bibitem [{\citenamefont {Luding}(2008)}]{Luding2008}%
  \BibitemOpen
  \bibfield  {author} {\bibinfo {author} {\bibfnamefont {S.}~\bibnamefont
  {Luding}},\ }\href {\doibase 10.3166/ejece.12.785-826} {\bibfield  {journal}
  {\bibinfo  {journal} {Revue europ\'{e}enne de g\'{e}nie civil}\ }\textbf
  {\bibinfo {volume} {12}},\ \bibinfo {pages} {785} (\bibinfo {year}
  {2008})}\BibitemShut {NoStop}%
\bibitem [{\citenamefont {Bini}\ and\ \citenamefont {Jones}(2007)}]{Bini2007}%
  \BibitemOpen
  \bibfield  {author} {\bibinfo {author} {\bibfnamefont {M.}~\bibnamefont
  {Bini}}\ and\ \bibinfo {author} {\bibfnamefont {W.~P.}\ \bibnamefont
  {Jones}},\ }\href {\doibase 10.1063/1.2709706} {\bibfield  {journal}
  {\bibinfo  {journal} {Physics of Fluids}\ }\textbf {\bibinfo {volume} {19}},\
  \bibinfo {pages} {035104} (\bibinfo {year} {2007})}\BibitemShut {NoStop}%
\bibitem [{\citenamefont {Li}\ and\ \citenamefont {Kuipers}(2003)}]{Li2003}%
  \BibitemOpen
  \bibfield  {author} {\bibinfo {author} {\bibfnamefont {J.}~\bibnamefont
  {Li}}\ and\ \bibinfo {author} {\bibfnamefont {J.~A.~M.}\ \bibnamefont
  {Kuipers}},\ }\href {\doibase 10.1016/S0009-2509(02)00599-7} {\bibfield
  {journal} {\bibinfo  {journal} {Chemical Engineering Science}\ }\textbf
  {\bibinfo {volume} {58}},\ \bibinfo {pages} {711} (\bibinfo {year}
  {2003})}\BibitemShut {NoStop}%
\bibitem [{\citenamefont {Zhu}\ \emph {et~al.}(2007)\citenamefont {Zhu},
  \citenamefont {Zhou}, \citenamefont {Yang},\ and\ \citenamefont
  {Yu}}]{Zhu2007}%
  \BibitemOpen
  \bibfield  {author} {\bibinfo {author} {\bibfnamefont {H.}~\bibnamefont
  {Zhu}}, \bibinfo {author} {\bibfnamefont {Z.}~\bibnamefont {Zhou}}, \bibinfo
  {author} {\bibfnamefont {R.}~\bibnamefont {Yang}}, \ and\ \bibinfo {author}
  {\bibfnamefont {A.}~\bibnamefont {Yu}},\ }\href {\doibase
  10.1016/j.ces.2006.12.089} {\bibfield  {journal} {\bibinfo  {journal}
  {Chemical Engineering Science}\ }\textbf {\bibinfo {volume} {62}},\ \bibinfo
  {pages} {3378} (\bibinfo {year} {2007})}\BibitemShut {NoStop}%
\bibitem [{\citenamefont {Aidun}\ and\ \citenamefont
  {Clausen}(2010)}]{Aidun2010}%
  \BibitemOpen
  \bibfield  {author} {\bibinfo {author} {\bibfnamefont {C.~K.}\ \bibnamefont
  {Aidun}}\ and\ \bibinfo {author} {\bibfnamefont {J.~R.}\ \bibnamefont
  {Clausen}},\ }\href {\doibase 10.1146/annurev-fluid-121108-145519} {\bibfield
   {journal} {\bibinfo  {journal} {Annual Review of Fluid Mechanics}\ }\textbf
  {\bibinfo {volume} {42}},\ \bibinfo {pages} {439} (\bibinfo {year}
  {2010})}\BibitemShut {NoStop}%
\bibitem [{\citenamefont {Mendoza}\ \emph {et~al.}(2010)\citenamefont
  {Mendoza}, \citenamefont {Boghosian}, \citenamefont {Herrmann},\ and\
  \citenamefont {Succi}}]{Mendoza2010}%
  \BibitemOpen
  \bibfield  {author} {\bibinfo {author} {\bibfnamefont {M.}~\bibnamefont
  {Mendoza}}, \bibinfo {author} {\bibfnamefont {B.~M.}\ \bibnamefont
  {Boghosian}}, \bibinfo {author} {\bibfnamefont {H.~J.}\ \bibnamefont
  {Herrmann}}, \ and\ \bibinfo {author} {\bibfnamefont {S.}~\bibnamefont
  {Succi}},\ }\href {\doibase 10.1103/PhysRevLett.105.014502} {\bibfield
  {journal} {\bibinfo  {journal} {Physical Review Letters}\ }\textbf {\bibinfo
  {volume} {105}},\ \bibinfo {pages} {014502} (\bibinfo {year}
  {2010})}\BibitemShut {NoStop}%
\bibitem [{\citenamefont {Mendoza}\ \emph {et~al.}(2011)\citenamefont
  {Mendoza}, \citenamefont {Herrmann},\ and\ \citenamefont
  {Succi}}]{Mendoza2011}%
  \BibitemOpen
  \bibfield  {author} {\bibinfo {author} {\bibfnamefont {M.}~\bibnamefont
  {Mendoza}}, \bibinfo {author} {\bibfnamefont {H.~J.}\ \bibnamefont
  {Herrmann}}, \ and\ \bibinfo {author} {\bibfnamefont {S.}~\bibnamefont
  {Succi}},\ }\href {\doibase 10.1103/PhysRevLett.106.156601} {\bibfield
  {journal} {\bibinfo  {journal} {Physical Review Letters}\ }\textbf {\bibinfo
  {volume} {106}},\ \bibinfo {pages} {156601} (\bibinfo {year}
  {2011})}\BibitemShut {NoStop}%
\bibitem [{\citenamefont {Bhatnagar}\ \emph {et~al.}(1954)\citenamefont
  {Bhatnagar}, \citenamefont {Gross},\ and\ \citenamefont
  {Krook}}]{Bhatnagar1954}%
  \BibitemOpen
  \bibfield  {author} {\bibinfo {author} {\bibfnamefont {P.~L.}\ \bibnamefont
  {Bhatnagar}}, \bibinfo {author} {\bibfnamefont {E.~P.}\ \bibnamefont
  {Gross}}, \ and\ \bibinfo {author} {\bibfnamefont {M.}~\bibnamefont
  {Krook}},\ }\href {\doibase 10.1103/PhysRev.94.511} {\bibfield  {journal}
  {\bibinfo  {journal} {Physical Review Letters}\ }\textbf {\bibinfo {volume}
  {94}},\ \bibinfo {pages} {511} (\bibinfo {year} {1954})}\BibitemShut
  {NoStop}%
\bibitem [{\citenamefont {Guo}\ \emph {et~al.}(2002)\citenamefont {Guo},
  \citenamefont {Zheng},\ and\ \citenamefont {Shi}}]{Guo2002}%
  \BibitemOpen
  \bibfield  {author} {\bibinfo {author} {\bibfnamefont {Z.}~\bibnamefont
  {Guo}}, \bibinfo {author} {\bibfnamefont {C.}~\bibnamefont {Zheng}}, \ and\
  \bibinfo {author} {\bibfnamefont {B.}~\bibnamefont {Shi}},\ }\href {\doibase
  10.1103/PhysRevE.65.046308} {\bibfield  {journal} {\bibinfo  {journal}
  {Physical Review E}\ }\textbf {\bibinfo {volume} {65}},\ \bibinfo {pages}
  {046308} (\bibinfo {year} {2002})}\BibitemShut {NoStop}%
\bibitem [{\citenamefont {Eswaran}\ and\ \citenamefont
  {Pope}(1988)}]{Eswaran1988}%
  \BibitemOpen
  \bibfield  {author} {\bibinfo {author} {\bibfnamefont {V.}~\bibnamefont
  {Eswaran}}\ and\ \bibinfo {author} {\bibfnamefont {S.~B.}\ \bibnamefont
  {Pope}},\ }\href {\doibase 10.1016/0045-7930(88)90013-8} {\bibfield
  {journal} {\bibinfo  {journal} {Computers \& Fluids}\ }\textbf {\bibinfo
  {volume} {16}},\ \bibinfo {pages} {257} (\bibinfo {year} {1988})}\BibitemShut
  {NoStop}%
\bibitem [{\citenamefont {Elghobashi}(1994)}]{Elghobashi1994}%
  \BibitemOpen
  \bibfield  {author} {\bibinfo {author} {\bibfnamefont {S.}~\bibnamefont
  {Elghobashi}},\ }\href {\doibase 10.1007/BF00936835} {\bibfield  {journal}
  {\bibinfo  {journal} {Applied Scientific Research}\ }\textbf {\bibinfo
  {volume} {52}},\ \bibinfo {pages} {309} (\bibinfo {year} {1994})}\BibitemShut
  {NoStop}%
\bibitem [{\citenamefont {Nash}\ \emph {et~al.}(2008)\citenamefont {Nash},
  \citenamefont {Adhikari},\ and\ \citenamefont {Cates}}]{Nash2008}%
  \BibitemOpen
  \bibfield  {author} {\bibinfo {author} {\bibfnamefont {R.~W.}\ \bibnamefont
  {Nash}}, \bibinfo {author} {\bibfnamefont {R.}~\bibnamefont {Adhikari}}, \
  and\ \bibinfo {author} {\bibfnamefont {M.~E.}\ \bibnamefont {Cates}},\ }\href
  {\doibase 10.1103/PhysRevE.77.026709} {\bibfield  {journal} {\bibinfo
  {journal} {Physical Review E}\ }\textbf {\bibinfo {volume} {77}},\ \bibinfo
  {pages} {026709} (\bibinfo {year} {2008})}\BibitemShut {NoStop}%
\bibitem [{\citenamefont {Burton}\ and\ \citenamefont
  {Eaton}(2005)}]{Burton2005}%
  \BibitemOpen
  \bibfield  {author} {\bibinfo {author} {\bibfnamefont {T.~M.}\ \bibnamefont
  {Burton}}\ and\ \bibinfo {author} {\bibfnamefont {J.~K.}\ \bibnamefont
  {Eaton}},\ }\href {\doibase 10.1017/S0022112005006889} {\bibfield  {journal}
  {\bibinfo  {journal} {Journal of Fluid Mechanics}\ }\textbf {\bibinfo
  {volume} {545}},\ \bibinfo {pages} {67} (\bibinfo {year} {2005})}\BibitemShut
  {NoStop}%
\bibitem [{\citenamefont {Tanaka}\ and\ \citenamefont
  {Eaton}(2010)}]{Tanaka2010}%
  \BibitemOpen
  \bibfield  {author} {\bibinfo {author} {\bibfnamefont {T.}~\bibnamefont
  {Tanaka}}\ and\ \bibinfo {author} {\bibfnamefont {J.~K.}\ \bibnamefont
  {Eaton}},\ }\href {\doibase 10.1017/S0022112009992023} {\bibfield  {journal}
  {\bibinfo  {journal} {Journal of Fluid Mechanics}\ }\textbf {\bibinfo
  {volume} {643}},\ \bibinfo {pages} {177} (\bibinfo {year}
  {2010})}\BibitemShut {NoStop}%
\end{thebibliography}%

\end{document}